      [1999/12/01 v1.4c Il Nuovo Cimento]
\documentclass{cimento}

\usepackage{graphicx} 
\usepackage{epsfig} 
\title{K Rare Decays with NA62}
\author{E.~Marinova\from{ins:x}
\thanks{Speaker, on behalf of NA48/2 and NA62 collaborations.}}
\instlist{\inst{ins:x} INFN Sezione di Perugia,
       Perugia.
                           }
\PACSes{\PACSit{13.25.Eb}{Decays of K mesons}
\PACSit{11.30.Hv}{Flavor symmetries}
\PACSit{12.15.Mn}{Neutral currents}
}
\begin{document}
\makeatletter
\newcommand{\rmnum}[1]{\romannumeral #1}
\newcommand{\Rmnum}[1]{\expandafter\@slowromancap\romannumeral #1@}
\makeatother

\maketitle

\begin{abstract}

The long-term goal of NA62 is to measure the ultra rare $K^{\pm}\rightarrow \pi^{\pm} \nu\bar{ \nu}$  decay with a sensitivity of $10^{-12}$ per event. This is done by using the decay - in - flight technique which allows a signal acceptance of $\sim10\%$ . The aim is to collect about 100 signal events in two years of data taking with a background to signal ratio smaller than 10\%. The principle of the experimental measurement and the layout of the detector are presented.
During 2007/2008 a dedicated run devoted to NA62 prototype tests and study of $K_{e2}$ decays was taken. The first phase of the NA62 experiment is aiming at a high precision test of the lepton universality by measuring the helicity suppressed ratio $R_{K}$. The preliminary result based on 40\% of the 2007 NA62 data sample,  $R_K = K_{e2}/K_{\mu2} = (2.500\pm0.016)\times10^{-5}$, which is the first result with a precision better than 1\%, is consistent with the Standard Model.

Aiming at charge asymmetry measurements, the NA48/2 experiment collected an unprecedented amount of charged $K_{3\pi}$ events. The large samples allowed a precision measurement of rare charged kaon decays. New measurements of the $K^\pm_{\pi\ell\ell}$ decays based on the full NA48/2 data sample collected during 2003/2004 are reported in this paper. Samples of about 7200 reconstructed $K^{\pm} \rightarrow \pi^\pm e^+e^-$ events, and more than 3000 $K^{\pm} \rightarrow \pi^\pm\mu^+\mu^-$ events, with a few percent background contamination, have been collected.
A precise measurement of the branching fractions and the form factors of the rare decays $K^{\pm}_{\pi\ell\ell}$  were performed. 
Measurements of the CP-violating and the forward-backward  asymmetries are reported.
\end{abstract}

\section{Introduction}

The field of rare kaon decays gives opportunities for various interesting studies, like testing low energy structure of QCD by studying the long distance effects dominated  $K^\pm \rightarrow\pi^\pm \ell^+ \ell^-$ decays, testing the lepton universality in $K_{\ell2}$ decays, and searching for New Physics (NP) effects in the rarest decays $K^{\pm}\rightarrow \pi^{\pm} \nu\bar{ \nu}$ .

The NA48 experiments have a long history in studying direct CP violation effects in the kaon system, but many rare kaon decays were studied as well. The NA62 experiment is a continuation of the CERN based kaon research program which probes the very rare decays and searches for NP contributions.

\section{The $K^{\pm}\rightarrow \pi^{\pm} \nu\bar{ \nu}$ decay with NA62}

$K^{\pm}\rightarrow \pi^{\pm} \nu\bar{ \nu}$ is a flavour-changing neutral current (FCNC) process that proceeds via one loop Z -penguin and W -box diagrams. Due to the t-quark contribution in the loops, it is sensitive to the $V_{td}$ parameter of the CKM matrix. Theoretically, it is one of the cleanest kaon decays, which are most sensitive to NP because the hadronic matrix element can be extracted with a very high precision from $K_{\ell3}$ decays~\cite{ref:sm}. 
The calculated values for its branching ratio BR$(K^{\pm}\rightarrow \pi^{\pm} \nu\bar{ \nu})\times 10^{10}$ vary between 0.75 up to 4:
the Standard Model (SM) prediction is 0.85$\pm$ 0.07~\cite{ref:sm}, the Minimal Flavour Violation Model
 -- 1.91~\cite{ref:mvf},
the Enhanced Electroweak Penguin model -- 
0.75 $\pm$ 0.21~\cite{ref:eewp},
the Extra Down type Singlet Quark model --
up to 1.5~\cite{ref:edsq}, and
the Minimal Supersymmetric Standard Model (MSSM)
up to 4.0~\cite{ref:mssm}. The only experimental measurement, based on 7 events, done by the E787/E949 experiment, gives BR$(K^{\pm}\rightarrow \pi^{\pm} \nu\bar{ \nu})\times 10^{10} = 1.73^{+ 1.15}_ {-1.05}$~\cite{ref:exp}. The central value is two times larger than the SM calculation but, due to the large experimental errors, the result is perfectly in agreement with the theory.

$K^{\pm}\rightarrow \pi^{\pm} \nu\bar{ \nu}$ decay has one charged track signature in the initial state, one charged track in the final state, and nothing else. All other kaon decays, as well as accidentals, with a similar signature are backgrounds. The successful background rejection relies on (1) precise timing for associating the secondary $\pi^+$ to the decaying $K^+$; (2) kinematic rejection of two and three-body kaon decays; (3) $\mu$ and $\gamma$ vetoing; (4) particle identification(ID) for $K^+/\pi^+$ and $\pi/\mu$ separation.
\begin{figure}
\label{fig:na62det}
\begin{center}
\resizebox*{.90\textwidth}{!}{\includegraphics{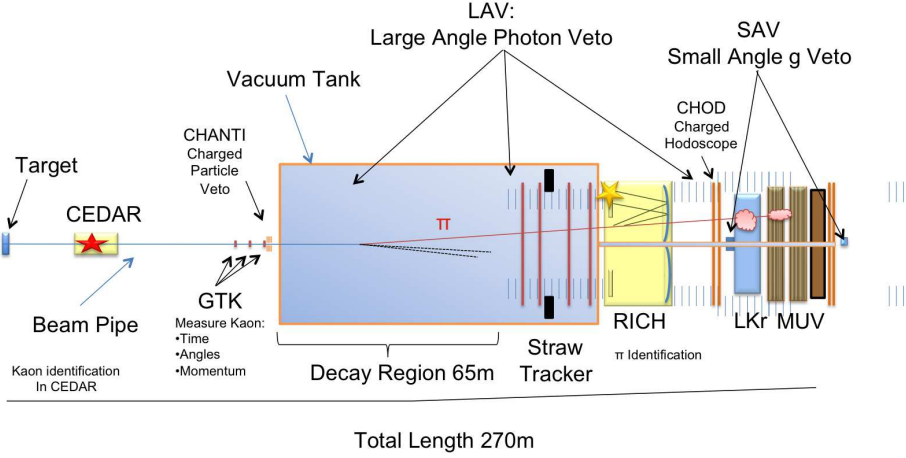}}
\end{center}
\caption{The NA62 detector.}
\end{figure}

 Currently, the experiment is in the construction phase. It will reuse the existing infrastructure for the NA48 experiments at CERN.
 In the proposed setup, 400 GeV/c primary protons are delivered by CERN SPS on a Be target.  A 75 GeV/c, narrow momentum-band (1.2\% RMS) hadron beam, out of which $\sim$ 6 \% are the $K^+$, is formed. To reduce the accidentals background, a positive identification of the kaons is provided by a differential Cherenkov counter (CEDAR)(see fig.~\ref{fig:na62det}). The challenge for this detector is to identify the K in a $\sim$ 800 MHz beam rate environment.
The position of the kaons, their time and momenta are measured by 3 silicon pixel detectors forming the beam spectrometer (Gigatracker) which will operate in vacuum . Each station covers an area of $60 \times 27$ mm$^2$ with pixels size $300 \times 300\mu$m$^2$
wide and 200$\mu$m thick. The time resolution for each station should be better than 200 ps in order to provide proper reconstruction of the K tracks.

The decay region is 65 m long and it starts 5 m after the last Gigatracker station. The momenta and the coordinates of the charged secondary decay products will be  measured by a spectrometer built of a dipole magnet and four chambers of straw tubes, each chamber of which has 4 layers of straws rotated by 45 degrees with respect to each other. 
To minimize the multiple scattering of the outgoing pion, the magnetic spectrometer will operate in vacuum upstream the decay region. This means that the decay and spectrometer regions are not separated and share a common vacuum volume.  A set of ring anticounters, CHANTI, surrounding the last Gigatracker station form a Òguard ringÓ and veto charged particles coming from the collimator. 

An 18 m long Neon filled Ring Imaging Cherenkov Detector (RICH) at atmospheric pressure  is placed between the forth Straw Tracker Chamber and the Charged Hodoscope (CHOD). It is used for a $\pi$ - $\mu$ separation for particles with momenta 15 - 35 GeV/c and to measure the time of the passing particles. Its inefficiency for the given momentum range is lower than $10^{-2}$ and its time resolution is better than 100 ps~\cite{rich}. The precise timing of the charged $\pi$ is obtained combining information from the RICH and the CHOD.

The photon veto system of the experiment, consisting of several detectors, ensures the hermetic coverage for photons flying at angles between 0 and  $\sim$50 $\mu$rad originating from $K^+$ decays in the fiducial region.
A high resolution electromagnetic liquid krypton calorimeter, LKr, used in the NA48 experiments,  detects and measures the energies of the electromagnetic showers. Another veto ring - Intermediate Ring Calorimeter (IRC) - is placed in front of the LKr and covers the inactive LKr region around the beam pipe. The LKr, together with IRC, are used for vetoing  photons flying at angles from 1 $\mu$rad to 8.5 $\mu$rad. 
12 Large Angle Photon Veto stations (LAV), surrounding the decay and detector volumes, are built of radial arrays of lead-glass blocks, arranged in overlapping layers.  the LAV system covers up to 50 $\mu$rad. The photon veto system is completed by a Small Angle shashlyk Calorimeter (SAC) placed at the end of the beamline. It's purpose is to provide a hermetic coverage for photons flying at small angles  below 1 $\mu$rad.

The hadron and muon ID downstream of the LKr is done by the muon veto detector (MUV) which consists of two parts - a hadron calorimeter and x- and y- segmented scintilator planes separated by an 80 cm thick iron wall. The MUV is followed by a dipole magnet that deflects the charged particles out of the acceptance of the SAC.

\begin{figure}
\label{fig:regions}
\begin{center}
\resizebox*{.45\textwidth}{!}{\includegraphics{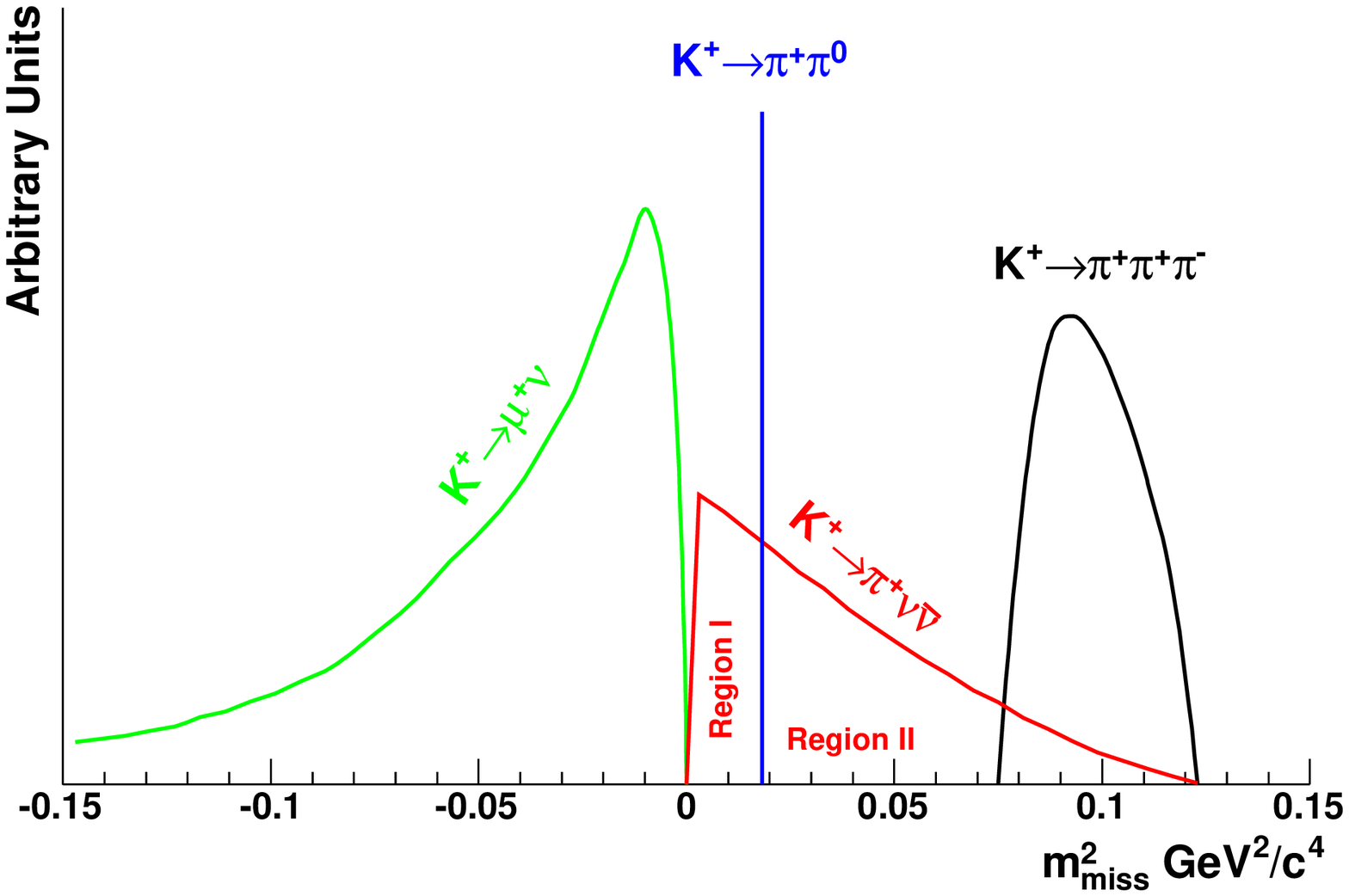}}
\put(-20,100){a)}
\resizebox*{.45\textwidth}{!}{\includegraphics{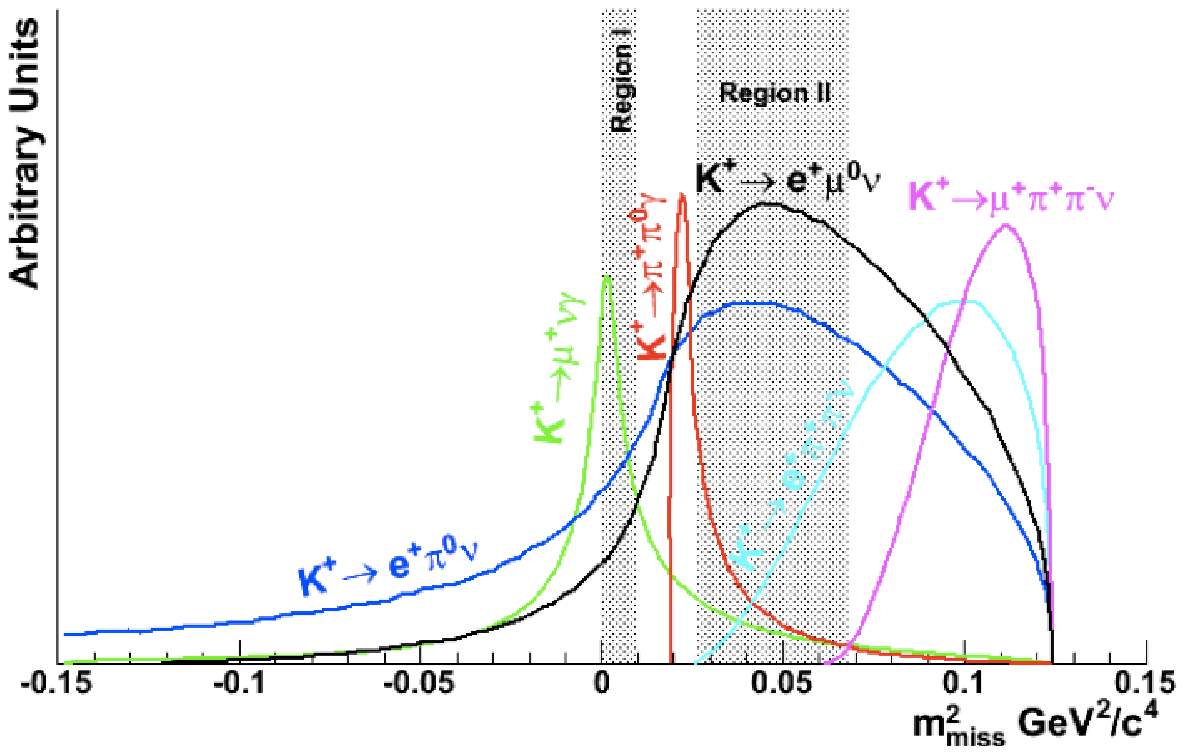}}
\put(-20,100){b)}
\end{center}
\caption{{\bf{a)}} $m^2_{miss}$ distribution for $K^{\pm}\rightarrow \pi^{\pm} \nu\bar{ \nu}$ signal, and kinematically constrained background; {\bf{b)}}  and kinematically unconstrained background (Monte Carlo simulation based).}
\end{figure}

The sensitivity evaluation, with the configuration described in the proposal~\cite{proposal}, is based on  the missing square mass studies $m^2_{miss} = (P_K - P_{\pi})^2$, assuming that the outcoming particle is a $\pi$. Using  $m^2_{miss}$, two types of background can be distinguished: one kinematically constrained corresponding to $\sim$ 92 \% of the kaon decays (see fig.~\ref{fig:regions}), and kinematically unconstrained (see fig.~\ref{fig:regions}{\bf{b}}). In fig.~\ref{fig:regions}{\bf{a}} one could see that the $K^{\pm}\rightarrow\pi^{\pm}\pi^0~(K_{2\pi})$ decays are particularly dangerous because they are in the middle of the  $K^{\pm}\rightarrow \pi^{\pm} \nu\bar{ \nu}$ signal region. Therefore, two independent $K_{2\pi}$ "free" regions on both sides of the $K_{2\pi} $ missing mass peak, called region \Rmnum{1} and region \Rmnum{2} , are chosen for the analysis. The kinematical rejection power is estimated to be at the order of $10^{-5}$.
The $K_{2\pi} $ background can be additionally suppressed if there is a cut on the charged $\pi$ momentum to be lower than 35 GeV/c. Then, a 40 GeV/c $\pi^0$ is difficult to be missed by the photon veto system. The inefficiency of the photon veto system is estimated to be between 2.0 and 3.5 $\times 10^{-8}$.

The MUV assures additional rejection for $K_{\mu2}$  background and it's muon rejection inefficiency is at the order of $10^{-5}$. This information, combined with the kinematical rejection and the particle ID from the RICH (with $ < 10^{-2}$ misidentification inefficiency for $p_{\pi} < 35$ GeV/c) gives a total suppression inefficiency of the  $K_{\mu2}$ background as low as $10^{-12}.$
For all the kinematically unconstrained background, coming from kaon decays with much smaller BR, the background rejection strongly relies on the photon vetoes, muon veto and particle ID. 

According to the SM calculation, in one year($\sim$100 days/year;$\sim$60\% efficiency) it is expected to collect $\sim$ 55 events.
Preliminary sensitivities studies show that the targets of 10\% of signal acceptance and 10\% background appear to be within reach. 
NA62 is optimized for a $K^{\pm}\rightarrow \pi^{\pm} \nu\bar{ \nu}$ analysis but given the high demands for the performance of the detectors, the experiment can support a broader physics program 
\footnote[1]{The new possibilities were discussed at The NA62 physics workshop; the web pages can be found at http://indico.cern.ch/conferenceDisplay.py?confId=65927}.

\section{$K^{\pm}\rightarrow \pi^{\pm}\ell^+\ell^-$ decays with NA48/2}

The NA48/2 fixed target experiment at CERN-SPS uses simultaneous $K^+$ and $K^-$ beams with momenta of $60\pm3$ GeV/c momenta. 
The experiment was designed for charge asymmetry measurements of $K_{3\pi}$ decays, and along with the main $K_{3\pi}$ samples, it has collected large amounts of charged K rare decays.  The main components of the detector are a magnetic spectrometer for measuring the momenta of the charged particles, a hodoscope providing  fast time signals for triggering, LKr measuring the energies of the electromagnetic showers, and a muon detector for $\mu$ ID.
A more detailed description of the detector system can be found elsewhere~\cite{na48det}. The NA48/2 experiment has recently finalized the measurement of the FCNC transitions decays $K^{\pm}\rightarrow \pi^{\pm}\ell^+\ell^-$ for both the $e$ and the $\mu$ mode.


The rare decays $K^{\pm}\rightarrow \pi^{\pm}\ell^+\ell^-$  induced at the one-loop level in the SM constitute a ground for testing the SM and its low-energy extensions. The decay rate for these decays is given in terms of the phase space factor and the form-factors which depend on the main kinematical variable $z = M_{\ell \ell}^2/ M_{K}^2$, where $M_{\ell \ell}$ is the dilepton mass, and $M_{K}$ is the kaon mass.
The spectrum in the dilepton invariant mass is then given by
$\frac{d\Gamma}{dz}=\frac{\alpha^2M_K^2}{12\pi(4\pi)^4}\lambda^{3/2}(1,z,r_{\pi}^2)\sqrt{1-4\frac{r_{\ell}^2}{z}}\left(1+2\frac{r_{\ell}^2}{z}\right)|W(z)|^2,$
with $r_{\ell}=m_{\ell}/M_K$, \\$4r_{\ell}^2\leq z \leq
(1-r_{\pi})^2$~\cite{chpt98} and $\lambda(a,b,c)=a^2+b^2+c^2-2ab-2ac-2bc$.

The form factor $W(z)$ is calculated in next--to--leading order (NLO)~\cite{chpt98}
The following parameterizations of the form factors W(z) are considered :
(1) Linear: $W(z)=G_FM_K^2|f_0|(1+\lambda z)$ with normalization
$|f_0|$ and linear slope $\lambda$;
(2) ChPT at NLO: $W(z)=G_FM_K^2W_+^{pol}+W_+^{\pi\pi}(z)$,~\cite{chpt98}, with free parameters $a_+$ and $b_+ $, entering the polynomial term of the equation;
(3) Combined framework of ChPT and large-Nc QCD: $ W(z)\equiv W(\tilde{w},\beta,z)$,~\cite{friot}, with free parameters $\tilde{w}$ and $\beta$;
(4) The ChPT parametrization involving the resonances $a$ and $\rho$ contribution
$ W(z)\equiv W(M_{a},M_{\rho},z) $~\cite{pervushin}, with resonance masses
$(M_{a}, M_{\rho})$ treated as free parameters.
 Each pair of parameters can be measured, and then, a model dependent BR can be calculated for each pair. In addition, a model independent ratio can be performed in the accessible kinematical range of $z$.

The interference of the long-distance $K \rightarrow \pi\gamma^*$
amplitude and the short-distance contribution leads to an asymmetry
between the widths of $K^+\rightarrow \pi^+e^+e^-$ and
$K^-\rightarrow \pi^-e^+e^-$ which is a clear signal of direct $CP$
violation. This quantity is defined by
$\Delta(K^{\pm}_{\pi^{\pm}\ell^+\ell^-}) = \frac{\Gamma(K^+\rightarrow\pi^+e^+e^-)-\Gamma(K^-\rightarrow\pi^-e^+e^-)}{\Gamma(K^+\rightarrow\pi^+e^+e^-)+\Gamma(K^-\rightarrow\pi^-e^+e^-)}\sim
\Im m~\lambda_t,$
where $\lambda_t=V_{td}V_{ts}^*$ ~\cite{chpt98}. However, with
$\Im m~ \lambda_t \sim 10^{-4}$~\cite{cpvpiee}, it is very
difficult to detect this effect within the SM.

\subsection{The rare decay $K^{\pm}\rightarrow \pi^{\pm}e^+e^-$}
\begin{figure}
\begin{center}
\resizebox*{.33\textwidth}{!}{\includegraphics{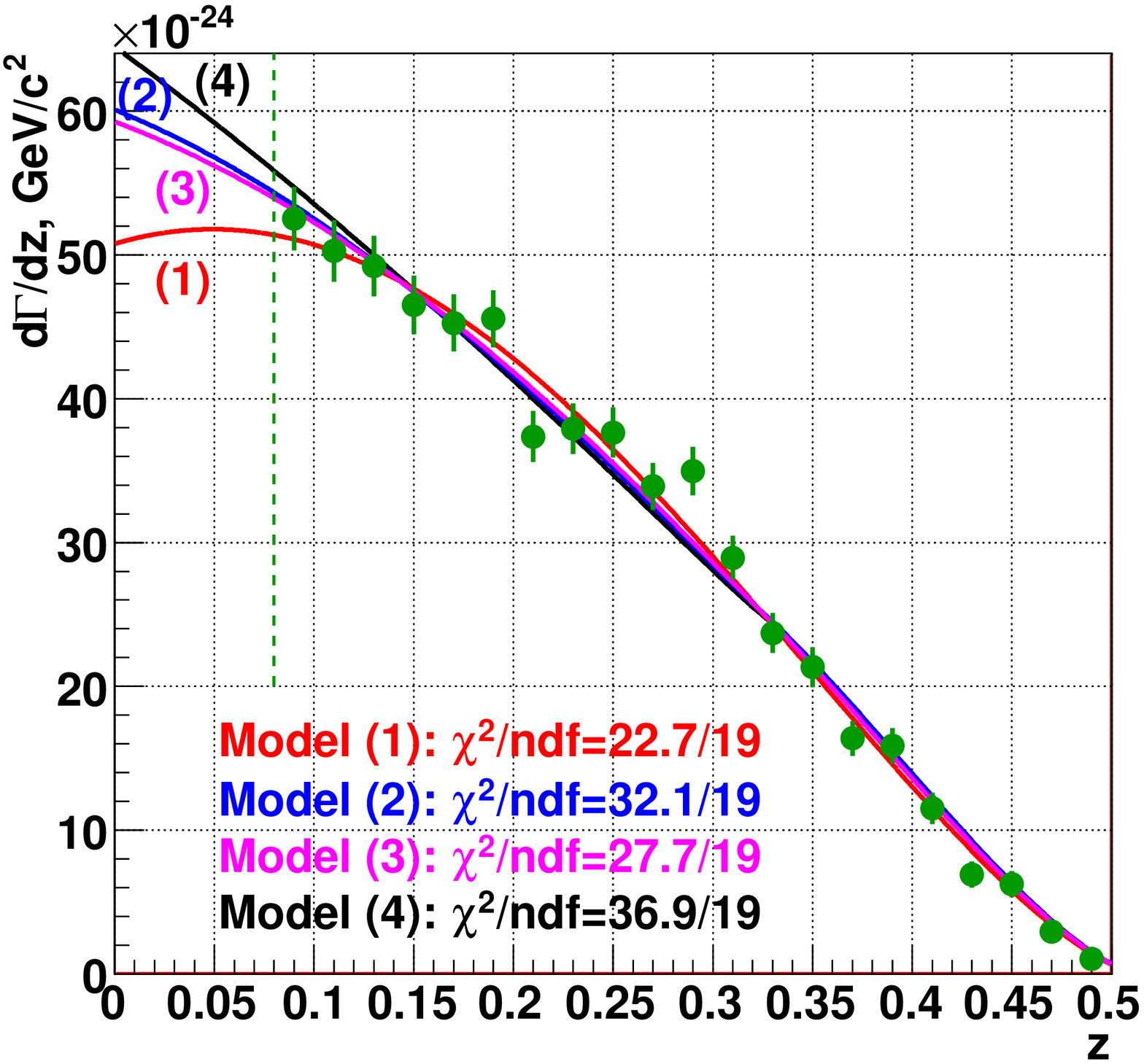}}
\put(-20,100){a)}
\resizebox*{.33\textwidth}{!}{\includegraphics{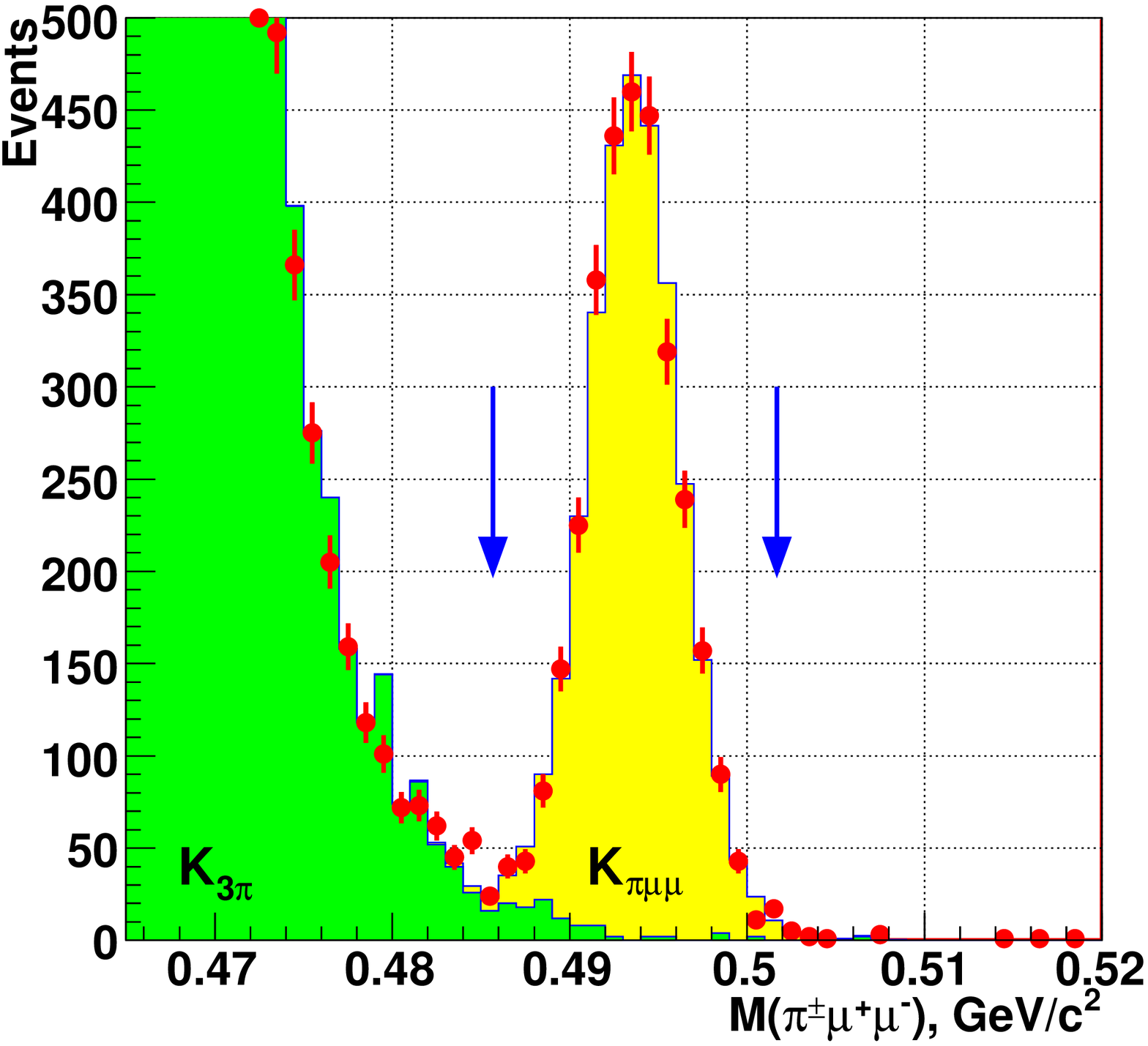}}
\put(-20,100){b)}
\resizebox*{.33\textwidth}{!}{\includegraphics{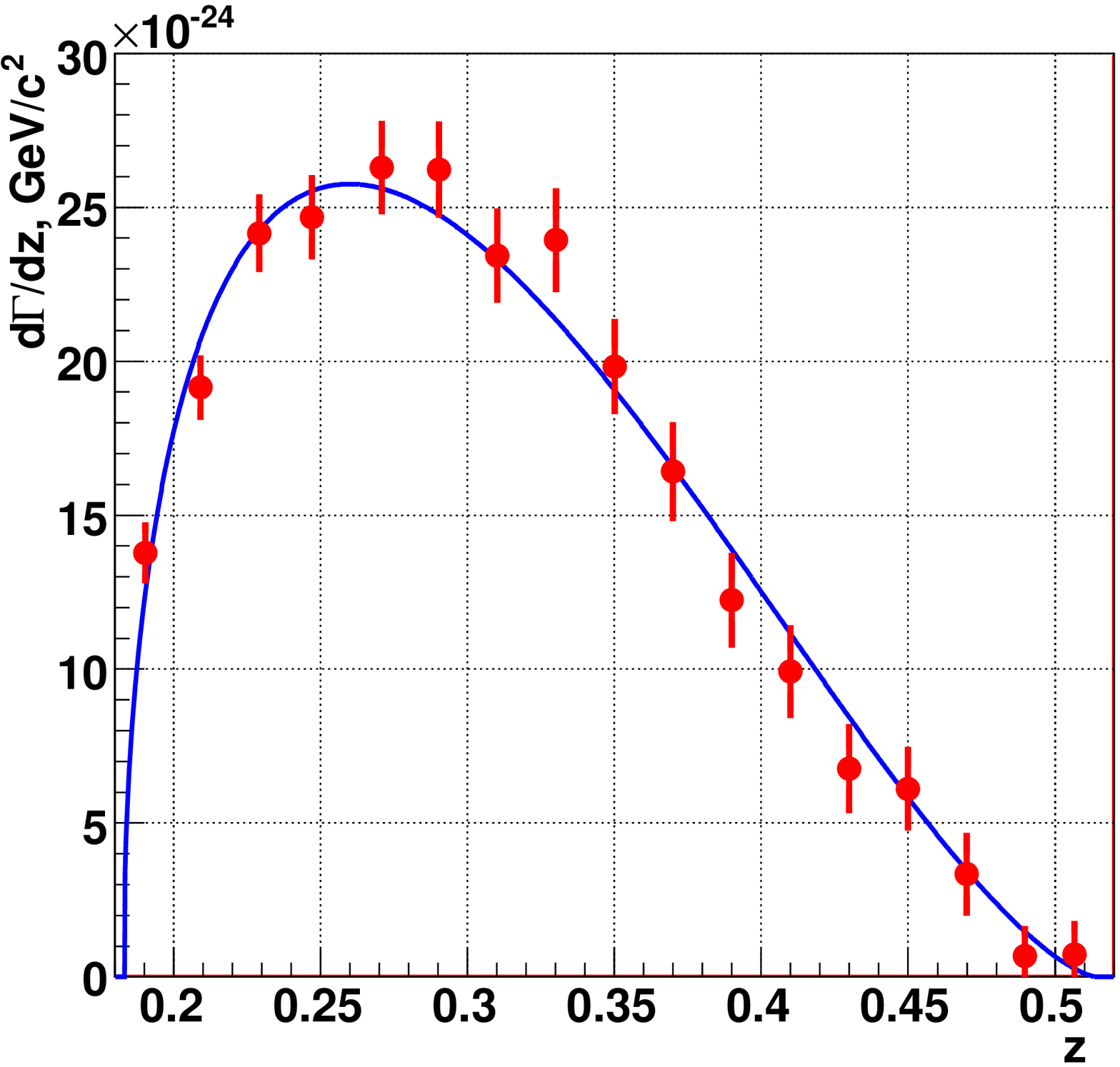}}
\put(-20,100){c)}
\end{center}
\caption{
{\bf{a)}} The reconstructed $d\Gamma_{\pi ee}/dz$ spectrum fitted to the four models for the form factors. {\bf{b)}} The signal region for $K^{\pm}\rightarrow \pi^{\pm}\mu^+\mu^-$ events (in yellow), and the $K_{3\pi} $background tail (in green). {\bf{c)}} The reconstructed $d\Gamma_{\pi\mu\mu}/dz$ spectrum fitted to a linear form factor.
}
\label{fig:piee}
\end{figure}

The measurement of the  $K^{\pm}\rightarrow \pi^{\pm}e^+e^-$ decay is based on 7253 events, with a background of $(1.0 \pm 0.1)\%$. The very similar decay $K^\pm\rightarrow\pi^\pm\pi^0_D$, where $\pi^0\rightarrow e^+e^-\gamma$, was chosen as a normalization channel.
The accessible kinematical region in $z$ is above $z<0.08$ due to the presence of background coming from the normalization channel which cannot be efficiently suppressed\footnote[2]{There is the possibility to remeasure it with NA62 whose hermetic veto system will allow suppressing backgrounds with additional photons, and access the region $z<0.08$. }. The reconstructed $d\Gamma_{K_{\pi ee}}/dz$ spectrum was fitted to the four models, and the form factor parameters were extracted. The four models cannot be distinguished in the visible kinematical region for $K^{\pm}\rightarrow \pi^{\pm}e^+e^-$. However, below $z<0.08$, the theory predicts different behavior of the four models. The form factor fits to the $d\Gamma_{K_{\pi ee}}/dz$ spectrum are presented in fig.~\ref{fig:piee} {\bf{a}}, and the results are reported in table~\ref{tab:fitresults}, together with the model independent BR in the visible kinematic region, and the combined result of the four models for the BR over the whole $z$ range. The results of the first three models and the BR are in agreement with the results reported in~\cite{prevexp1}, \cite{prevexp2},\cite{prevexp3}, and with the theoretical prediction for $a_+= -0.6^{+0.3}_ {-0.6}$~\cite{prades}.  Model 4 was never tested before.

The first measurement of the CP violating asymmetry, done by NA48/2, \\$\Delta(K^\pm_{\pi^{\pm}e^+e^-}) = (-2.2\pm1.5_{stat}\pm0.6_{syst})\times10^{-2}$ is consistent with no CP violation. However, its precision is far from the SM expectation~\cite{cpvpiee}.
\subsection{The rare decay $K^{\pm}\rightarrow \pi^{\pm}\mu^+\mu^-$}
\begin{table}
  \caption{Results of fits to the four models and the BR of $K^{\pm}\rightarrow \pi^{\pm}\ell^+\ell^-$ decays.}
  \label{tab:fitresults}
  \begin{tabular}{c | c | c | c}
    \hline

Model & Parameter & Results & Results \\
&  & $K^{\pm}\rightarrow \pi^{\pm}e^+e^-$ & $K^{\pm}\rightarrow \pi^{\pm}\mu^+\mu^-$ \\
\hline
 &$ \lambda$ &  2.32 $\pm$ 0.18& 3.11 $\pm$ 0.56\\
Model 1  & $|f_0|$ &  $0.531 \pm 0.016$ &0.470 $\pm$ 0.039\\
\hline
 & $a_+$ & -0.578 $\pm$ 0.016 &-0.575 $\pm$ ±0.038\\
Model 2  & $b_+$ &  -0.779 $\pm$ 0.066&-0.813 $\pm$ 0.142 \\
\hline
&$\tilde{w}$& 0.057 $\pm$ 0.007 &0.064 $\pm$ 0.014\\
Model 3 &$\beta$&0.531 $\pm$ 0.016&0.064 $\pm$ 0.014\\
\hline
&$M_a$&0.974 $\pm$ 0.035&1.014 $\pm$ 0.090\\\
 Model 4 &$M_\rho$ & 0.716 $\pm$ 0.014&0.725 $\pm$ 0.028\\
\hline
Combined result &BR  & $(3.11 \pm 0.12) \times 10^{-7}$&--\\
\hline
Model independent &BRmi & $z > 0.08$& full range\\
 & & (2.28 $\pm$ 0.08) $\times 10^{-7}$ & $(9.25 \pm 0.62)\times10^{-8}$\\
\hline
  \end{tabular}
\end{table}

The analysis is based on 3120 reconstructed events, 4 times more than the total world's sample, with a background of $(3.3\pm0.5)\%$. The main technique of background estimation is based on choosing events with two $\mu$ with the same sign from the data sample, and the result is confirmed by a $K_{3\pi}$ MC simulation. 
Each of the four models for the form factors provides a reasonable fit to the data: the values of $\chi^2$ per degree of freedom are 12.0/15, 14.8/15, 13.7/15 and 15.4/15. The results of the fits are reported in table~\ref{tab:fitresults}. The data sample size is insufficient to distinguish between the models considered.
A measurement of the CP violating asymmetry, $\Delta(K^{\pm}_{\pi^{\pm}\mu^+\mu^-}) = (1.1\pm2.3)\times10^{-2}$, is consistent with CP conservation, but its precision is far from the theoretical predictions~\cite{cpvpiee}.
Another interesting observable, the forward-backward asymmetry in terms of the
$\Theta_{K\mu}$ angle between three-momenta of the kaon and the muon of opposite sign in the $\mu^+\mu^-$
rest frame, was measured for the first time:
$A_{FB} = \frac{(N(\Theta_{K\mu}>0)-N(\Theta_{K\mu}<0))}{(N(\Theta_{K\mu}>0)+N(\Theta_{K\mu}<0))} = (-2.4\pm1.8)\times10^{-2}$, where the error is dominated by the statistical uncertainty.
The achieved precision does not reach the upper limits of the SM~\cite{afbsm} and the MSSM~\cite{afbmssm}, both at the order of $10^{-3}$.
The results on the BR agrees with two of the previous measurements~\cite{pmm:prevexp2}, \cite{pmm:prevexp3}, and disagrees with~\cite{pmm:prevexp1} .
The measurements on the form factors agree with the $K^{\pm}\rightarrow \pi^{\pm}e^+e^-$ results of NA48/2~\cite{piee}, with the $\lambda$ value measured by~\cite{pmm:prevexp2},  and with
theoretical expectation of $a_+= -0.6^{+0.3}_ {-0.6}$~\cite{prades}.

\section{The $R_K = K_{e2(\gamma)}/K_{\mu2(\gamma)}$ measurement - NA62, phase \Rmnum{1} }

\begin{figure}
\label{fig:rk}
\begin{center}
\resizebox*{0.33\textwidth}{!}{\includegraphics{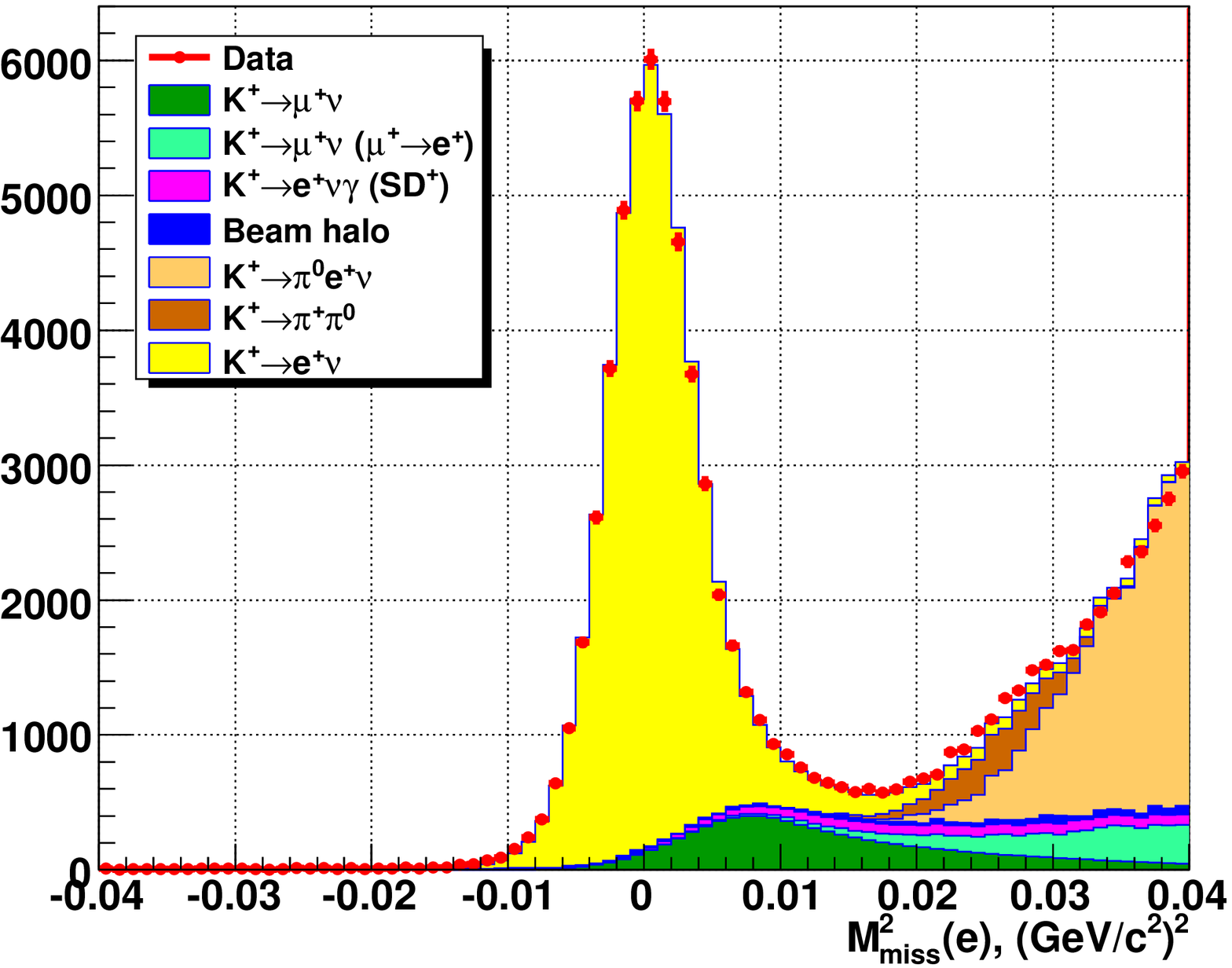}}
\put(-20,90){a)}
\resizebox*{0.33\textwidth}{!}{\includegraphics{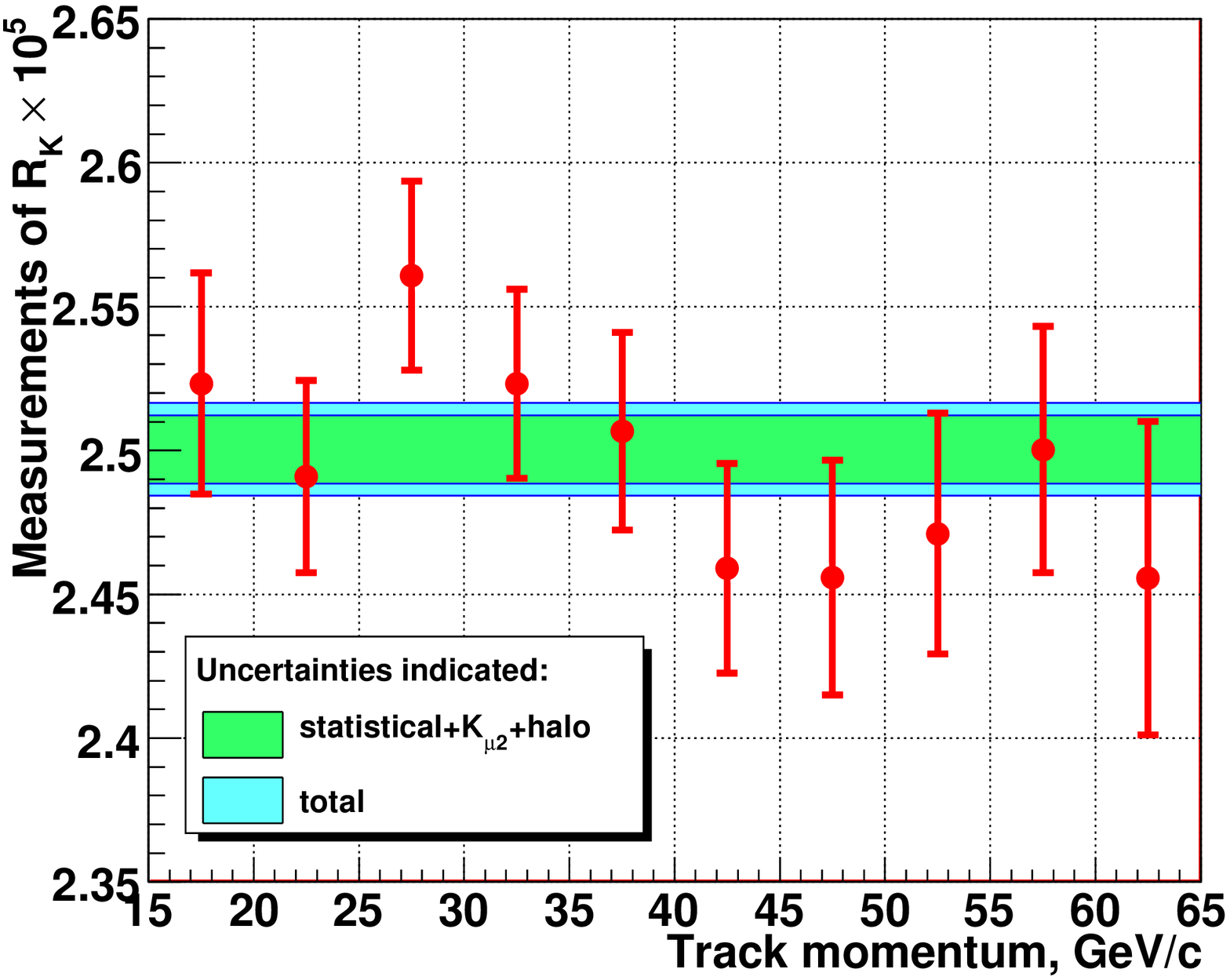}}
\put(-20,90){b)}
\resizebox*{0.33\textwidth}{!}{\includegraphics{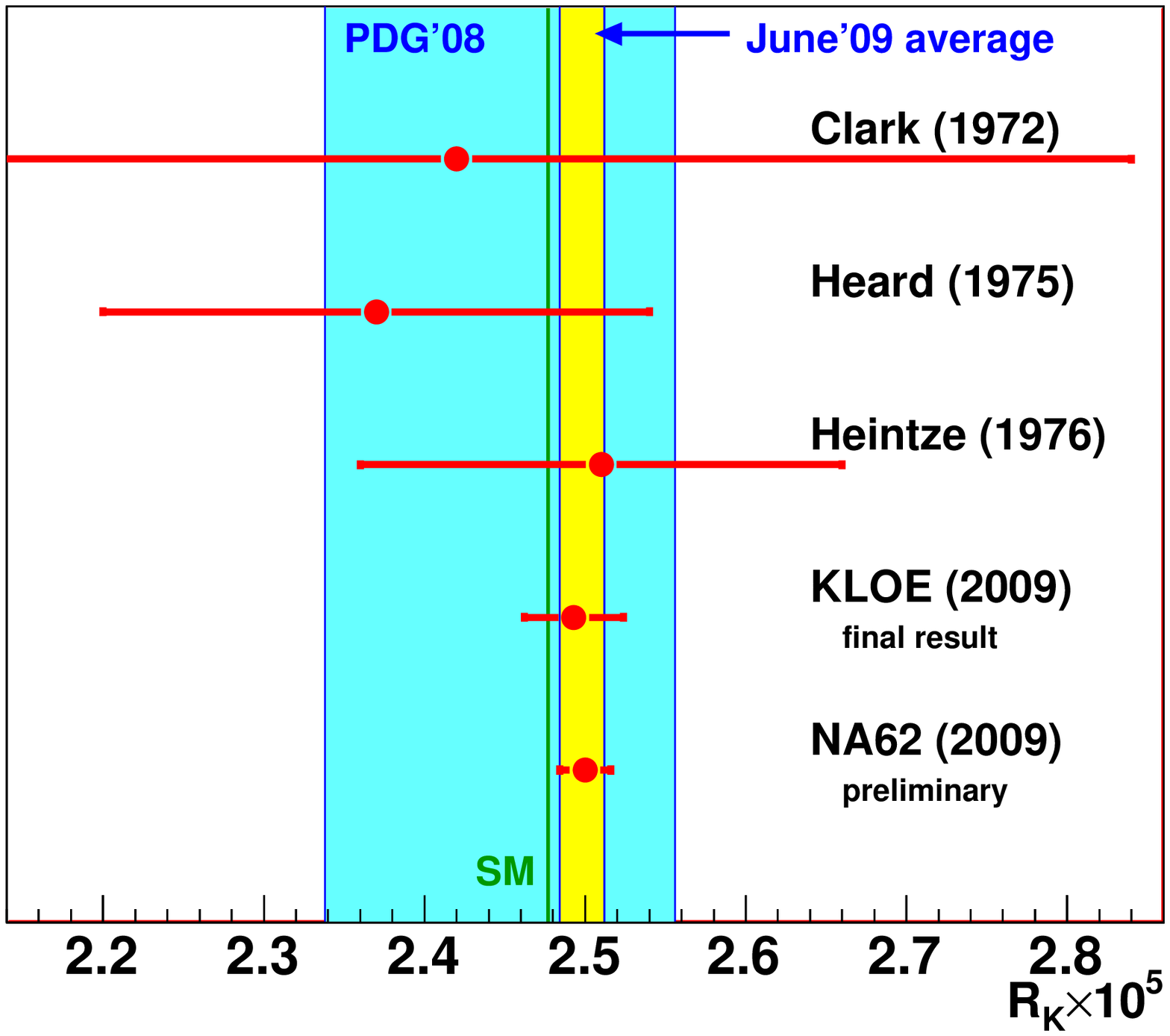}}
\put(-20,90){c)}
\end{center}
\caption{ {\bf{a)}} Reconstructed squared missing mass distribution $M^2_{\mathrm{miss}}$($e$ hypothesis) for the $K_{e2}$ events and the different backgrounds contributions. {\bf{b)}}  Measurements of $R_K$ in independent lepton momentum bins.  {\bf{c)}} $R_K$ experimental results and SM expectation.}
\end{figure}

The leptonic ratio $R_K = \left(\frac{m_e}{m_{\mu}}\right)^2\left(\frac{m_K^2-m_e^2}{m_K^2-m_{\mu}^2}\right)^2 (1+\delta R_{QED}) = (2.477\pm0.001)\times10^{-5}$(SM)~\cite{cirigliano}, where $\delta R_{QED} = (-3.78 \pm 0.04)\%$ is a correction due to the IB radiative process, can be predicted with a high accuracy within SM due to the cancellation of the hadronic uncertainties. The factor $\left(\frac{m_e}{m_{\mu}}\right)^2$ accounts for the helicity suppression of the $K_{e2}$ decay with respect to $K_{\mu2}$ decay. This helicity suppression enhances the sensitivity to non-SM lepton flavour violating (LFV) effects which are not ruled out experimentally. It was realized recently that LFV effects can shift the $R_K$ value by a few percent~\cite{masiero1},\cite{masiero2}.

The worlds average on $R_K$ in 2008 is $R_K=(2.45\pm0.11)\times10^{-5} , (\delta R_K/R_K=4.5\%)$,~\cite{pdg08} based on three experiments in the 1970s. KLOE has recently published their improved precision result, $R_K=(2.493\pm0.031)\times10^{-5},  (\delta R_K/R_K=1.3\%)$ based on 13800 $K_{e2}$ candidates, with 16\% background~\cite{kloe}.
NA62 set a goal of collecting 150000 $K_{e2}$ events with a background less than 10\% for measuring $R_K$ with a 0.5\% precision.
For this measurement, NA62 used the old NA48/2 experimental setup, slightly optimized for the $K_{e2}$ measurement. For example, the kaon momentum was increased to $74\pm2$ GeV/c. The narrow momentum band aimed to minimize the contribution of the momentum resolution in the kinematical variables.

The preliminary results reported here are on 40\% of the total statistics which corresponds to 51 089 $K_{e2}$ events with a background of $(8.03\pm0.23)\%$ (see fig.~\ref{fig:rk}{\bf{a}}). The estimate is to reach 135 thousand events with the full sample.$15.56 \times 10^6$ events were collected for the normalization $K_{\mu2}$ channel, with a very low background of 0.25\%. The analysis was done in independent lepton momentum bins (see fig.~\ref{fig:rk}{\bf{b}}). The preliminary NA62 result of $R_K$ has a much better precision than the previous measurements: $R_K= (2.5 \pm 0.012_{stat} \pm 0.011_{syst})\times 10^{-5}, (\delta R_K/R_K = 0.64\%)$(see fig.~\ref{fig:rk}{\bf{c}}), where the statistical uncertainties dominate the the systematical ones. The main systematical uncertainties to this measurement are due to the  $K_{\mu2}$  background contribution to $K_{e2}$ with a $\mu$ misidentified as an electron, beam halo background estimation, the positron ID, the simulation of the IB radiative process, the trigger dead time and uncertainties due to corrections on the geometric acceptance.
The analysis of the full data sample should achive the desired precision of 0.5\%.

\section{Conclusions}

Using a sample of 7253 events  $K^\pm \rightarrow \pi^\pm e^+e^-$ collected by NA48/2 during 2003/2004, a high precision measurement of the form factors within four different models, and the BR have been measured. A first measurement of the CP-violating asymmetry, $\Delta(K^{\pm}_{\pi^{\pm}e^+e^-}) = (-2.2\pm1.5_{stat}\pm0.6_{syst})\times10^{-2}$, is consistent with no CP violation. Using an unprecedented amount of  $K^\pm \rightarrow \pi^\pm \mu^+\mu^-$ decays, with a very low background of $(3.3\pm0.5)\%$, a model independent BR has been measured $BR = (9.62 \pm 0.62) \times 10^{-8}$, and the form factors described within the four available models were extracted. An improved precision measurement of the CP violating charge asymmetry and (for the first time) the forward-backward asymmetry of the integrated decay rate were presented.

NA62 phase \Rmnum{1} data taking in 2007/08 was dedicated to a precision measurement of $R_K$. 
The analysis of a partial Ke2 sample ($\sim$40\%) reached the record accuracy of 0.64\% and the preliminary result: $R_K= (2.500\pm0.016)\times10^-5$
is compatible with the SM prediction. NA62 phase \Rmnum{2} is a challenging experiment aiming at measuring $K^\pm\rightarrow\pi^\pm\nu\bar{\nu}$ decay by collecting O(100) events with $B/S<10\%$. It has been approved by CERN SPSC and Research Board. Data taking is foreseen to start in 2012.

\end{document}